# Ultra-broadband Heteronuclear Hartmann-Hahn Polarization Transfer

Xijia Miao

Center for Magnetic Resonance Research, Minnesota University Medical School, 2021 Sixth Street, SE, Minneapolis, Minnesota, MN55455, USA.

E-mail: miao@cmrr.umn.edu

**Abstract**

It is showed on the basis of the multiple-quantum operator algebra space formalism that ultra broadband heteronuclear Hartmann-Hahn polarization transfer could be achieved by the amplitude- and frequency-modulation quasi-adiabatic excitation (90 degree) pulses, while it is usually difficult for the adiabatic inversion pulses to achieve effectively broadband Hartmann-Hahn transfer in a heteronuclear coupled two-spin system. The adiabatic inversion and quasi-adiabatic excitation pulses have an important property that within their activation bandwidth flip angle of the pulses is independent of the pulse duration and the bandwidth increases as the pulse duration. This property is important for the construction of the heteronuclear Hartmann-Hahn transfer sequence with the quasi-adiabatic 90 degree pulses. Theoretical analysis and numerical simulation show that the heteronuclear Hartmann-Hahn transfer is performed in the even-order multiple-quantum operator subspace of the two-spin system. The multiple-quantum operator algebra space formalism may provide a powerful guide to the construction of ultra-broadband heteronuclear Hartmann-Hahn transfer sequences with the quasi-adiabatic 90 degree pulses.

## 1. Introduction

The Hartmann-Hahn polarization transfer experiment [1] is one of the most important nuclear magnetic resonance (NMR) experiments and has an extensive application in high-resolution NMR spectroscopy both in liquids and solids [1-6]. The Hartmann-Hahn transfer sequence may be used as a basic polarization-transfer building block to enhance the NMR signal intensity of dilute nuclei, which usually have a low spin polarization, and as a mixing sequence to achieve correction among different nuclear spins in molecules to help determination of molecular structures. In high-resolution NMR spectroscopy in liquid broadband heteronuclear Hartmann-Hahn transfer sequences usually were derived from heteronuclear decoupling sequences [7-11]. These decoupling sequences such as WALTZ, MLEV, and DIPSI families [7, 8] usually are of composite-pulse sequences of rectangular radiofrequency (RF) pulses or amplitude-modulation shaped RF pulses. Such Hartmann-Hahn transfer sequences have been used extensively in structural determination of organic molecules and large biomolecules [12, 13]. Usually these sequences need to dissipate much more RF power and have a high peak RF power in order that effective bandwidth to cover the full chemical-shift ranges of the nuclear spins in molecules can be obtained for the Hartmann-Hahn transfer. This generates significant resonance shifts due to sample heating which may cause problems for applications of the Hartmann-Hahn transfer sequences in high-resolution NMR spectroscopy. Therefore, a high-performance heteronuclear Hartmann-Hahn transfer sequence has been highly desired which has a lower peak RF power and dissipates low RF power but covers large chemical shift ranges. Having been stimulated by the success that the decoupling sequences based on adiabatic inversion pulses [14, 15] can achieve ultra-broadband heteronuclear decoupling in high-field NMR spectroscopy [16-18], several researchers have suggested to exploit adiabatic pulses to construct the Hartmann-Hahn sequences [19, 20]. Although a high tolerance to inhomogeneous RF field may be achieved for these sequences [19-23], it is really difficult to obtain a broadband Hartmann-Hahn transfer with these sequences with a low RF power. The adiabatic pulses have the superior performances [14, 15]: (a) this type of pulses have a much wider inversion bandwidth but dissipate much less RF power and have much lower peak RF power than the rectangular pulses; (b) the pulses have a high tolerance to inhomogeneous RF field. These advantageous performances should attribute to the fact that such pulses consist of pairs of amplitude- and frequency-modulation functions which fulfill the adiabatic condition [14, 15]. It has been desired highly that the superior performances of adiabatic pulses can be introduced into the Hartmann-Hahn transfer sequences by using the adiabatic pulses to build up the sequences. However, so far broadband heteronuclear Hartmann-Hahn sequences built up with adiabatic inversion pulses with a low RF power have not yet been found in high-resolution NMR spectroscopy in liquids. Other researchers suggested adiabatic Hartmann-Hahn transfer sequences with amplitude-modulation pulses, as be seen in Ref. [23-26]. These sequences usually are used in NMR spectroscopy in solids. It is expected that it is usually difficult for these sequences to obtain broadband Hartmann-Hahn transfer with a low RF power since these sequences use only amplitude-modulation RF field.

 It is well-known that flip angle of RF pulse is usually proportional to the pulsewidth for a hard pulse or more generally is dependent in a complex form on the pulsewidth for an amplitude-modulation shaped pulse [7]. Thus, for a given RF power an excitation pulse (90 degree) or an inversion pulse (180 degree) has a fixed pulsewidth [5, 7]. However, there are also another type of pulses whose flip angle is independent of their pulsewidth. An example is adiabatic inversion pulses and quasi-adiabatic excitation (90 degree) pulses [27], as can be seen below. The type of pulses has many advantageous performances and characteristic properties themselves. One of the important properties may be that the conversion efficiency of the initial longitudinal magnetization $M_0$ in a single spin system under the pulse is determined only by the constant adiabatic factor of the pulse within the conversion bandwidth, as investigated recently [27]. This indicates that flip angle of the pulse is determined only by the constant adiabatic factor. Furthermore, the flip angle is independent of the pulse duration within the conversion bandwidth, as investigated below, and the conversion bandwidth increases as the pulse duration and the square of the RF power [15, 27, 35, 36]. This property shows that adiabatic inversion pulses may not be proper to construct broadband heteronuclear Hartmann-Hahn transfer sequences according to the theoretic analysis and numerical simulation below. It is understanding this point that leads me to the new idea to construct heteronuclear Hartmann-Hahn transfer sequences by using the quasi-adiabatic excitation (90 degree) pulses. As investigated below, this property also shows that quasi-adiabatic 90 degree pulses could be suitable for constructing heteronuclear Hartmann-Hahn transfer sequences in contrast to the adiabatic inversion pulses. In this paper it is proposed to use the amplitude- and frequency-modulation quasi-adiabatic 90 degree pulses to build up heteronuclear Hartmann-Hahn transfer sequences. A theoretic analysis is presented for the possible mechanics of the heteronuclear Hartmann-hahn transfer based on the multiple-quantum operator algebra space formalism [28]. Numerical simulation confirms the theoretic analysis. With the help of the multiple-quantum operator algebra space formalism ultra-broadband heteronuclear Hartmann-Hahn transfer sequences could be constructed with the quasi-adiabatic 90 degree pulses.

## 2. The even-order multiple-quantum operator subspace

A heteronuclear coupled two-spin system under a pair of amplitude- and frequency- (or phase-) modulation radiofrequency (RF) pulses has the total spin Hamiltonian in the rotating frame when neglecting the relaxation effects:

$$H_T(t) = H_0(t) + H_1(t) \qquad (1)$$

where the time-dependent Hamiltonian $H_0(t)$ and $H_1(t)$ are the contributions from the pulses and the scalar interaction between the two spins, respectively,
$$H_0(t) = \Omega_i I_z + \omega_{1i}(t)(I_x \cos(\phi_i(t)) + I_y \sin(\phi_i(t))) \\ + \Omega_s S_z + \omega_{1s}(t)(S_x \cos(\phi_s(t)) + S_y \sin(\phi_s(t))) \quad (2)$$

$$H_1 = \pi J 2 I_z S_z, \quad (3)$$

where $\Omega_i$ and $\Omega_s$ are the chemical shifts of the spins I and S, respectively and J is the scalar coupling constant between the two spins; $\omega_{1q}(t)$ and $\phi_q(t)$ are the amplitude- and phase-modulation functions of the pulse applied to the spin $q$ ($q = i, s$), respectively. The total time-evolution propagator of the spin system under the pulses then can be written as

$$U_T(t) = T \exp[-i\int_0^t (H_0(t') + H_1(t'))dt'], \quad (4)$$

where $T$ is the Dyson time-ordering operator. This propagator can be decomposed into the product of the two factors corresponding to the two Hamiltonian operators $H_0(t)$ and $H_1(t)$, respectively [5],

$$U_T(t) = U_0(t)U_i(t) \quad (5)$$

where

$$U_0(t) = T \exp(-i\int_0^t H_0(t')dt'), \quad (6)$$

$$U_i(t) = T \exp(-i\int_0^t H_i(t')dt') \quad (7)$$

with the interaction Hamiltonian $H_i(t)$ in the interaction frame:

$$H_i(t) = U_0(t)^+ H_1(t) U_0(t). \quad (8)$$

Since the Hamiltonian $H_0(t)$ of Eq.(2) does not contain the interaction between the two spins the corresponding propagator $U_0(t)$ actually describes the time evolution of a non-interacting two-spin system under a pair of amplitude- and frequency-modulation pulses. Then, one has the following unitary transformations [29-32]:

$$U_0(t)^+ I_z U_0(t) = \alpha_i(\Omega_i,t)I_z + \beta_i(\Omega_i,t)I_x + \gamma_i(\Omega_i,t)I_y, \quad (9a)$$

$$U_0(t)^+ S_z U_0(t) = \alpha_s(\Omega_s,t)S_z + \beta_s(\Omega_s,t)S_x + \gamma_s(\Omega_s,t)S_y. \quad (9b)$$

This is due to the fact that each of the above unitary transformations is equivalent to the rotating transformation in the Lie algebra space *su(2)* of a single spin with magnetic quantum number I=1/2. Now by using the unitary transformations (9a) and (9b) the interaction Hamiltonian $H_i(t)$ of Eq.(8) is written as

$$H_i(t) = \pi J_{zz} 2I_z S_z + \pi J_{xx} 2I_x S_x + \pi J_{xy} 2I_x S_y + \pi J_{yx} 2I_y S_x + \pi J_{yy} 2I_y S_y \\ + \pi J_{zx} 2I_z S_x + \pi J_{zy} 2I_z S_y + \pi J_{xz} 2I_x S_z + \pi J_{yz} 2I_y S_z \quad (10)$$

where the coefficients $J_{pq} = J_{pq}(\Omega_i, \Omega_s, t)$ (p, q=x, y, z) are dependent on the rotating parameters $\alpha_p(\Omega_p,t)$, $\beta_p(\Omega_p,t)$, and $\gamma_p(\Omega_p,t)$ ($p = i, s$) in Eqs.(9a) and (9b). Obviously, the interaction Hamiltonian is dependent on the resonance offsets or the Larmor frequencies of the two spins: $H_i(t) = H_i(\Omega_i, \Omega_s, t)$. In Eq.(10) the first term is the longitudinal two-spin order operator ($2I_zS_z$), the last four terms are single-quantum coherence operators, and the rest four terms are even-order multiple-quantum (double- and zero-quantum) coherence operators [28]. It will be seen in next sections that for a pair of adiabatic inversion pulses which are applied simultaneously to the heteronuclear coupled two-spin system the dominating term in the interaction Hamiltonian $H_i(t)$ of Eq.(10) is the longitudinal two-spin order operator within their inversion band,

$$H_i(t) = \pi J_{zz} 2I_zS_z. \qquad (11)$$

where the effective coupling constant $J_{zz} \approx J$. It is well known that this Hamiltonian can not derive the Hartmann-Hahn transfer [5]. This may be the main reason why the adiabatic inversion pulses usually are not suitable for deriving the Hartmann-Hahn transfer. However, for the quasi-adiabatic excitation pulses (90 degree) the dominating terms in the interaction Hamiltonian $H_i(t)$ of Eq.(10) are the even-order multiple-quantum operators within their excitation band,

$$H_i(t) = \pi J_{xx} 2I_xS_x + \pi J_{xy} 2I_xS_y + \pi J_{yx} 2I_yS_x + \pi J_{yy} 2I_yS_y. \qquad (12)$$

This Hamiltonian plays an important role in deriving heteronuclear Hartmann-Hahn transfer. It also shows that the quasi-adiabatic excitation pulses could be useful for constructing broadband heteronuclear Hartmann-Hahn transfer sequences.

In the coupled two-spin system IS the even-order multiple-quantum operator subspace [28] contains the double-, and zero-quantum coherence operators, the longitudinal magnetization ($I_z$ and $S_z$) and two-spin order ($2I_zS_z$) operators. The conventional Hartmann-Hahn polarization transfer [1-6] usually is performed in the zero-quantum operator subspace that consists of the zero-quantum coherence operators and the longitudinal magnetization ($I_z$ and $S_z$) and two-spin order operators ($2I_zS_z$) in the two-spin system since the effective spin Hamiltonian to derive the transfer is a zero-quantum operator. However, the Hartmann-Hahn polarization transfer may be performed more generally in the even-order multiple-quantum operator subspace and the effective spin Hamiltonian to derive the transfer is an even-order multiple-quantum operator [28, 29]. One possible form for the even-order multiple-quantum Hamiltonian is given by Eq.(12) and can be generally written as

$$H_e(t) = H_{ZQ}(\Omega_i, \Omega_s, t) + H_{DQ}(\Omega_i, \Omega_s, t) \qquad (13)$$

where the zero- ($H_{ZQ}$) and double-quantum ($H_{DQ}$) operators are commutative with each other and are given by

$$H_{ZQ}(\Omega_i, \Omega_s, t) = \pi J_x^z (2I_xS_x + 2I_yS_y) + \pi J_y^z (2I_xS_y - 2I_yS_x), \qquad (14a)$$

$$H_{DQ}(\Omega_i,\Omega_s,t) = \pi J_x^D (2I_x S_x - 2I_y S_y) + \pi J_y^D (2I_x S_y + 2I_y S_x) \tag{14b}$$

with the parameters dependent on time and the resonance offsets:

$$J_x^Z = J_x^Z(\Omega_i,\Omega_s,t) = \frac{1}{2}(J_{xx} + J_{yy}), \quad J_y^Z = J_y^Z(\Omega_i,\Omega_s,t) = \frac{1}{2}(J_{xy} + J_{yx}),$$

$$J_x^D = J_x^D(\Omega_i,\Omega_s,t) = \frac{1}{2}(J_{xx} - J_{yy}), \quad J_y^D = J_y^D(\Omega_i,\Omega_s,t) = \frac{1}{2}(J_{xy} - J_{yx}).$$

Then the Hartmann-Hahn transfer derived by the even-order multiple-quantum Hamiltonian of Eq.(13) can be expressed as

$$U_e(t) I_z U_e^+(t) = S_z \tag{15}$$

where the even-order multiple-quantum unitary propagator is given by

$$U_e(t) = T \exp[-i \int_0^t H_e(t')dt']. \tag{16}$$

This propagator may be expressed generally in a unitary matrix form

$$U_e(t) = \begin{pmatrix} \alpha_{11} & 0 & 0 & \alpha_{14} \\ 0 & \alpha_{22} & \alpha_{23} & 0 \\ 0 & \alpha_{32} & \alpha_{33} & 0 \\ \alpha_{41} & 0 & 0 & \alpha_{44} \end{pmatrix}, \tag{17}$$

where the nonzero matrix elements are time-dependent and also dependent on the parameters $\{J_{pq}; p, q=x, y\}$ of Eqs.(14a) and (14b) and hence the resonance offsets. Now using the initial longitudinal magnetization:

$$I_z = \begin{pmatrix} \frac{1}{2} & 0 & 0 & 0 \\ 0 & \frac{1}{2} & 0 & 0 \\ 0 & 0 & -\frac{1}{2} & 0 \\ 0 & 0 & 0 & -\frac{1}{2} \end{pmatrix},$$

one can calculate the Hartmann-Hahn transfer of Eq.(15),

$$U_e(t) I_z U_e^+(t)$$

$$= \frac{1}{2}\begin{pmatrix} |\alpha_{11}|^2 - |\alpha_{14}|^2 & 0 & 0 & \alpha_{11}\alpha_{41}^* - \alpha_{14}\alpha_{44}^* \\ 0 & |\alpha_{22}|^2 - |\alpha_{23}|^2 & \alpha_{22}\alpha_{32}^* - \alpha_{23}\alpha_{33}^* & 0 \\ 0 & \alpha_{32}\alpha_{22}^* - \alpha_{33}\alpha_{23}^* & |\alpha_{32}|^2 - |\alpha_{33}|^2 & 0 \\ \alpha_{41}\alpha_{11}^* - \alpha_{44}\alpha_{14}^* & 0 & 0 & |\alpha_{41}|^2 - |\alpha_{44}|^2 \end{pmatrix}. \tag{18}$$

Equation (18) shows that the transformation $U_e(t) I_z U_e(t)^+$ may create generally double- and zero-quantum coherence, and the longitudinal magnetization and spin order operators. Obviously, this is due to the properties of the even-order multiple-quantum operators [28, 33, 34]. By comparing Eq.(18) with Eq.(15) one obtains the conditions for the complete Hartmann-Hahn transfer:

$$|\alpha_{11}|^2 - |\alpha_{14}|^2 = |\alpha_{32}|^2 - |\alpha_{33}|^2 = 1,$$

$$|\alpha_{22}|^2 - |\alpha_{23}|^2 = |\alpha_{41}|^2 - |\alpha_{44}|^2 = -1,$$

and $\quad \alpha_{11}\alpha_{41}^* - \alpha_{14}\alpha_{44}^* = 0,\ \alpha_{22}\alpha_{32}^* - \alpha_{23}\alpha_{33}^* = 0$.

If the spin Hamiltonian of Eq.(13) is time-independent, the even-order multiple-quantum propagator of Eq.(16) can be written generally as

$$U_e(t) = \exp[-i(f_i I_z + f_s S_z)]\exp[-i\pi t J_x^{*Z}(2I_x S_x + 2I_y S_y)]\exp[i(f_i I_z + f_s S_z)]$$
$$\times \exp[-i(g_i I_z + g_s S_z)]\exp[-i\pi t J_x^{*D}(2I_x S_x - 2I_y S_y)]\exp[i(g_i I_z + g_s S_z)]$$

(19)

where the parameters are given by

$$J_x^{*Z} = \sqrt{(J_x^Z)^2 + (J_y^Z)^2} \text{ and } \tan(f_i - f_s) = -J_y^Z / J_x^Z;$$
$$J_x^{*D} = \sqrt{(J_x^D)^2 + (J_y^D)^2} \text{ and } \tan(g_i + g_s) = J_y^D / J_x^D.$$

Using this propagator (19) one can calculate analytically the Hartmann-Hahn transfer:

$$U_e(t)I_z U_e(t)^+ = \alpha^i I_z + \gamma^i S_z$$
$$+ \beta_{xx}^i 2I_x S_x + \beta_{xy}^i 2I_x S_y + \beta_{yx}^i 2I_y S_x + \beta_{yy}^i 2I_y S_y$$

(20a)

and

$$U_e(t)S_z U_e(t)^+ = \alpha^s S_z + \gamma^s I_z$$
$$+ \beta_{xx}^s 2I_x S_x + \beta_{xy}^s 2I_x S_y + \beta_{yx}^s 2I_y S_x + \beta_{yy}^s 2I_y S_y$$

(20b)

where

$$\alpha^i = \alpha^s = \cos^2(\pi J_x^{*D} t)\cos^2(\pi J_x^{*Z} t) - \sin^2(\pi J_x^{*D} t)\sin^2(\pi J_x^{*Z} t),$$
$$\gamma^i = \gamma^s = \cos^2(\pi J_x^{*D} t)\sin^2(\pi J_x^{*Z} t) - \sin^2(\pi J_x^{*D} t)\cos^2(\pi J_x^{*Z} t),$$
$$\beta_{xx}^i = -\beta_{yy}^s = \sin(\pi J_x^{*D} t)\cos(\pi J_x^{*D} t)\sin(g_i + g_s)$$
$$\qquad + \sin(\pi J_x^{*Z} t)\cos(\pi J_x^{*Z} t)\sin(f_i - f_s),$$
$$\beta_{xy}^i = \beta_{yx}^s = -\sin(\pi J_x^{*D} t)\cos(\pi J_x^{*D} t)\cos(g_i + g_s)$$
$$\qquad + \sin(\pi J_x^{*Z} t)\cos(\pi J_x^{*Z} t)\cos(f_i - f_s),$$
$$\beta_{yx}^i = \beta_{xy}^s = -\sin(\pi J_x^{*D} t)\cos(\pi J_x^{*D} t)\cos(g_i + g_s)$$
$$\qquad - \sin(\pi J_x^{*Z} t)\cos(\pi J_x^{*Z} t)\cos(f_i - f_s),$$
$$\beta_{yy}^i = -\beta_{xx}^s = -\sin(\pi J_x^{*D} t)\cos(\pi J_x^{*D} t)\sin(g_i + g_s)$$
$$\qquad + \sin(\pi J_x^{*Z} t)\cos(\pi J_x^{*Z} t)\sin(f_i - f_s).$$

The first two terms in Eqs.(20a) and (20b) are the longitudinal magnetization operators and the last four terms are double- and zero-quantum coherence operators. Therefore, the complete Hartmann-Hahn transfer ($I_z \to S_z$ or $S_z \to I_z$) occurs when the coefficient $\gamma^i = 1$ or $\gamma^s = 1$. The transformations of Eqs.(20a) and (20b) show that there are not only zero-quantum but also double-quantum coherences to be generated during the Hartmann-Hahn transfer period besides the longitudinal magnetization $I_z$ and $S_z$, indicating that the Hartmann-Hahn transfer derived by the even-order multiple-quantum Hamiltonian of Eq.(13) is performed in the even-order multiple-quantum operator subspace of the two-spin system (IS).

## 3. The amplitude- and frequency-modulation adiabatic inversion and quasi-adiabatic excitation pulses

The adiabatic inversion and quasi-adiabatic excitation pulses are composed of pairs of amplitude- and frequency-modulation functions. There are many methods to construct an adiabatic inversion pulse [14, 15, 27, 35, 36]. Recently, a general analytical method [27] has been proposed to construct high-performance adiabatic inversion and quasi-adiabatic 90 degree pulses. This method emphasizes the effect of adiabatic factor on the performance of adiabatic and quasi-adiabatic pulses, that is, the adiabatic factor can play an important role in the construction of the pulses. Here the adiabatic factor $P(\Delta\omega, t)$ is defined as [27]

$$P(\Delta\omega, t) = \frac{\omega_1(t)(d\Delta\omega(t)/dt) - \Delta\omega(t)(d\omega_1(t)/dt)}{(\omega_1(t)^2 + \Delta\omega(t)^2)^{3/2}}. \tag{21}$$

Particularly note that this definition of adiabatic factor is inversely that one in Refs.[14, 15, 35, 36]. Based on the general analytical method [27] a simple and convenient formula to design an adiabatic pulse is obtained,

$$\left|d\omega_r(t)/dt\right| + \frac{2\sqrt{3}}{9}\left|d\omega_1(t)/dt\right| = p\omega_1(t)^2, \quad (0 < p \ll 1) \tag{22}$$

where $0 \le t \le t_p$ and $t_p$ is the pulsewidth of the adiabatic pulse, p is the constant adiabatic factor, $\omega_1(t)$ and $\omega_r(t)$ are amplitude- and frequency-modulation functions, respectively. The phase-modulation function $\phi(t)$ is related to the frequency-modulation function by $\omega_r(t) = d\phi(t)/dt$. Now given any amplitude-modulation function $\omega_1(t)$ one can build up the corresponding frequency-modulation function $\omega_r(t)$ by the general analytical method [27] so that the adiabatic condition $P(\Delta\omega, t) \ll 1$ is always met over the whole chemical shift ($\Omega$) range or the whole resonance offset ($\Delta\omega = \Delta\omega(t) = \Omega - \omega_r(t)$): $|\Omega| < +\infty$ or $|\Delta\omega| < +\infty$. The second term in Eq.(22) is the contribution of time derivative of the amplitude-modulation function in the construction of the frequency-modulation function. If this term is small and can be neglected equation (22) is reduced to the conventional approximated formula [35, 36] to construct an adiabatic inversion pulse:

$$\left|d\omega_r(t)/dt\right| = p\omega_1(t)^2. \tag{23}$$

Equations (22) and (23) provide simple and convenient methods to construct adiabatic inversion and quasi-adiabatic 90 degree pulses. The more general method can be seen in Ref. [27]. It is particularly important that for the construction of an adiabatic inversion pulse with Eqs.(22) and (23) the constant adiabatic factor is set *p=1/3*, however, for a quasi-adiabatic 90 degree pulse the constant adiabatic factor is set *p=2.3*. The quasi-adiabatic 90 degree pulse is not an adiabatic pulse since it does not fulfill the adiabatic condition ($P(\Delta\omega,t) \ll 1$), but it has some similar properties of an adiabatic inversion pulse and is constructed with the same methods such as Eqs.(22) and (23) as the adiabatic inversion pulse [27]. Therefore it is called the quasi-adiabatic pulse here. The frequency-modulation function is approximately proportional to the product $p\omega_0^2$ of the adiabatic factor *p* and the square of the peak power ($\omega_0$) of the RF pulse [35, 36], as can be seen in Eqs.(22) and (23). Therefore, for some pulses such as the hyperbolic secant adiabatic pulse one could change the peak power of the pulse to obtain 90 or 180 degree flip angle although the adiabatic factor is not set p=2.3 or 1/3 when constructing the frequency-modulation function of the pulses by Eqs.(22) and (23) [15]. However, it is optimal for the construction of the adiabatic inversion and quasi-adiabatic excitation pulses using Eqs.(22) and (23) or more generally using the general analytical method [27] with settings p=1/3 and 2.3, respectively. Since the operating bandwidth for an adiabatic pulse and a quasi-adiabatic pulse is proportional approximately to the adiabatic factor *p* the excitation bandwidth of the quasi-adiabatic 90 degree pulse is about seven times wider than that of the adiabatic inversion pulse with the same RF power and pulsewidth.

That the quasi-adiabatic 90 degree pulses could be suitable for constructing heteronuclear Hartmann-Hahn transfer sequences is due to the fact that the pulses have the extraordinary property that flip angle of the pulses is independent of the pulse duration within the operating band of the pulses. It is well known that flip angle of the conventional rectangular pulses and the amplitude-modulation shaped pulses is usually dependent on the pulsewidth. Actually, flip angle of the adiabatic and quasi-adiabatic pulses is determined only by the constant adiabatic factor [27]. Numerical simulation shows that the adiabatic and quasi-adiabatic pulses have the important property that their flip angle is independent of the pulewidth. As a typical example, the conventional hyperbolic secant (backward-half part) quasi-adiabatic 90 degree pulse is investigated in the numerical simulation. For simplicity, the frequency-modulation function of the pulse is generated by Eq.(23) with the constant adiabatic factor *p=2.3* by starting the backward-half hyperbolic secant amplitude-modulation function $\omega_1(t) = \omega_0 \text{sech}(\beta t)$ (t ≥ 0). Figure 1 shows that bandwidth of the quasi-adiabatic excitation pulse increases as the pulse duration, but all the flip angles are the same (90 degree) for different pulse duration, as indicated by $M_z/M_0 \approx 0$, showing that the flip angle of the pulse does not change as the pulse duration. The adiabatic inversion pulses also have the same property as the quasi-adiabatic pulses. The flip angle of the pulses does not change as the pulse duration, as indicated by $M_z/M_0 \approx -1$ within the inversion bandwidth of the pulses in Figure 2, where the inversion profiles of the backward-half hyperbolic secant adiabatic inversion pulse with different pulse duration are plotted. However, the adiabatic inversion (180 degree) pulses are just not proper to derive the broadband Hartmann-Hahn transfer. This will be discussed in detailed below.

**(a) The adiabatic inversion pulses**

If the pair of pulses applied to the two heteronuclear spins are of adiabatic inversion pulses, then the conversion coefficients $\alpha_q(\Delta\omega_q, t_p)$ ($q=i, s$), etc., in Eqs.(9a) and (9b) should satisfy the following conditions.
(1) As shown in schematic Figure 3 (A), within the inversion band $\Delta W_q(t_p)$: $-W'_q(t_p) \leq \Delta\omega_q \leq W_q(t_p)$ ($\Delta\omega_q$ is the resonance offset of the spin $q$ ($q=i, s$) and $t_p$ is pulse duration), the coefficients in Eqs.(9a) and (9b) should fulfill the relationship:

$$\alpha_q(\Delta\omega_q, t_p) = -1, \; \beta_q(\Delta\omega_q, t_p) = 0, \text{ and } \gamma_q(\Delta\omega_q, t_p) = 0. \tag{24}$$

This shows that the initial longitudinal magnetization $M_0$ ($I_z$ or $S_z$) is inverted completely within the inversion band and in the duration $t_p$ of the pulses.
(2) in the transition regions, i.e., $W_q(t_p) < \Delta\omega_q < W_q(t_p) + \delta W_q(t_p)$ or $-W'_q(t_p) > \Delta\omega_q > -(W'_q(t_p) + \delta W'_q(t_p))$, the initial longitudinal magnetization $M_0$ is converted partly into the transverse magnetization $M_x$ and $M_y$,

$$\alpha_q(\Delta\omega_q, t_p) \neq 0, \; \beta_q(\Delta\omega_q, t_p) \neq 0, \text{ and } \gamma_q(\Delta\omega_q, t_p) \neq 0. \tag{25}$$

(3) in the large resonance offset ranges, i.e., $\Delta\omega_q \geq W_q(t_p) + \delta W_q(t_p)$ or $\Delta\omega_q \leq -(W'_q(t_p) + \delta W'_q(t_p))$, the initial longitudinal magnetization $M_0$ keeps unchanged,

$$\alpha_q(\Delta\omega_q, t_p) = 1, \; \beta_q(\Delta\omega_q, t_p) = 0, \text{ and } \gamma_q(\Delta\omega_q, t_p) = 0. \tag{26}$$

This indicates that the adiabatic inversion pulses do not take into action when the resonance offset is outside the inversion bandwidth.

It must be emphasized that the conversion coefficients $\alpha_q(\Delta\omega_q, t)$ ($q=i, s$) is independent of the pulse duration t ($t > t_p$) within the inversion band $\Delta W_q(t_p)$ for the adiabatic pulses, as shown in Figures 2 and 3 (A).

The average Hamiltonian theory [5] may explain approximately why the heteronuclear Hartmann-Hahn transfer may not be achieved by the adiabatic inversion pulses. For simplicity, examine the Hartmann-Hahn polarization transfer on the resonance offset plane region $SW(t_p)$: $-W'_q(t_p) \leq \Delta\omega_q \leq W_q(t_p)$ ($q=i, s$). Assume that a pair of simultaneous adiabatic inversion pulses are applied to the two-spin system. These pulses have the same inversion bandwidth $\Delta W_s(t_p) = \Delta W_i(t_p)$ which is dependent on the duration $t_p$ of the pulses, as shown in Figure 3. Obviously, $\Delta W_q(t) \geq \Delta W_q(t_p)$ if $t \geq t_p$ and $\Delta W_q(t) \leq \Delta W_q(t_p)$ if $t \leq t_p$ since the inversion bandwidth increases as the pulse duration for the adiabatic pulses and the flip angle is independent of the pulse duration, as shown in Figure 2. For convenience, the total duration of the Hartmann-Hahn transfer derived by the adiabatic inversion pulses is denoted by $T_p$ and here assume that $T_p \gg t_p$. The interaction Hamiltonian $H_i(t)$ to derive the transfer during the period $T_p$ is given generally by Eq.(10). The zero-order average Hamiltonian for the interaction Hamiltonian is written as [5],

$$\overline{H}_i^{(0)} = \frac{1}{T_p} \int_0^{T_p} H_i(t') dt'. \tag{27}$$

This integral can be expressed as the discrete sum:

$$\overline{H}_i^{(0)} = \frac{1}{n} \sum_{k=1}^{n} H_i(t_k) \tag{28}$$

where the number $n$ takes an enough large number and $(k-1)T_p/n \leq t_k \leq kT_p/n$. It should be noted that the interaction Hamiltonian $H_i(t)$ of Eq.(10) is generally dependent on the resonance offsets of the two spins: $H_i(t) = H_i(\Delta\omega_i, \Delta\omega_s, t)$. Now in the case that the pulse duration $t$ (or the Hartmann-Hahn transfer duration) is smaller than $t_p$ the interaction Hamiltonian $H_i(t)$ (t < t$_p$) is given by Eq.(10) within the inversion plane region (denoted by SW(t$_p$)): $-W'_q(t_p) \leq \Delta\omega_q \leq W_q(t_p)$ ($q=i, s$), as shown in Figure 3 (B), since on the region all the coefficients in Eq.(10) may be nonzero due to the condition that $\alpha_q(\Delta\omega_q, t) \neq 0$, $\beta_q(\Delta\omega_q, t) \neq 0$, and $\gamma_q(\Delta\omega_q, t) \neq 0$ ($q = i, s; t < t_p$). This is related to the fact that there are transition regions in the inversion profiles of the adiabatic pulses with pulse duration $t < t_p$ within the inversion band $\Delta W_q(t_p)$, as can be seen in Figure 3 (A). The interaction Hamiltonian may contain zero-, single- and double-quantum coherence and longitudinal two-spin order operators. Such a Hamiltonian may not derive generally the broadband Hartmann-Hahn transfer. However, when the pulse duration $t$ is longer than $t_p$ the interaction Hamiltonian $H_i(t)$ ($t > t_p$) of Eq.(10) within the inversion plane region SW(t$_p$) should be simple and contain only the longitudinal two-spin order operator ($2I_zS_z$) since in Eqs.(9a) and (9b) only the coefficients $\alpha_q(\Delta\omega_q, t) = -1$ ($q = i, s; t > t_p$) and any other coefficients equal zero within the inversion band $\Delta W_q(t_p)$. Then the zero-order average Hamiltonian of Eq.(28) within the inversion plane region SW(t$_p$) can be divided into two parts:

$$\overline{H}_i^{(0)} = \frac{1}{n}\sum_{t_k<t_p}H_i(t_k) + \frac{1}{n}\sum_{t_k\geq t_p}H_i(t_k).$$

The first part takes generally the form of Eq.(10), but the second part is the longitudinal two-spin order operator ($2I_zS_z$). Since the total duration $T_p$ of the Hartmann-Hahn transfer is much longer than the time $t_p$, that is, $T_p \gg t_p$, the second part is the dominating term and the first part can be neglected. Consequently the zero-order average Hamiltonian is approximately equal to the longitudinal two-spin order operator within the inversion plane region SW($t_p$), $\overline{H}_i^{(0)} \approx \pi J 2I_zS_z$. Obviously, this Hamiltonian can not derive the Hartmann-Hahn transfer within the inversion plane region SW($t_p$).

More generally, the time evolution propagator of Eq.(7) can be expressed as the sequence with sufficiently small intervals $\{\Delta t_k = T_p/n\}$:

$$U_i(T_p) = \exp(-iH_i(t_n + \tfrac{1}{2}\Delta t_n)\Delta t_n)\exp(-iH_i(t_{n-1} + \tfrac{1}{2}\Delta t_{n-1})\Delta t_{n-1}) \\ .....\exp(-iH_i(t_k + \tfrac{1}{2}\Delta t_k)\Delta t_k).....\exp(-iH_i(t_1 + \tfrac{1}{2}\Delta t_1)\Delta t_1) \qquad (29)$$

The total duration of the Hartmann-Hahn sequence can be divided into the two periods so that the sequence of Eqs.(29) can be expressed as the product of two parts. One part is the one with the evolution time $t$ shorter than $t_p$ and another is the rest part with evolution time $t$ longer than $t_p$,

$$U_i(T_p) = U_i(T_p, t_p)U_i(t_p, 0) \qquad (30)$$

where the propagators are defined as Eq.(7): $U_i(t, t_0) = T\exp(-i\int_{t_0}^{t} H_i(t')dt')$.

All these unitary propagators in Eq.(30) are usually dependent on the resonance offsets of the two spins. Since the interaction Hamiltonian $H_i(t)$ ($t < t_p$) of the unitary propagator $U_i(t_p, 0)$ is generally given by Eq.(10) within the region SW($t_p$), any initial magnetization $\rho(0)$ may be transferred into all the possible operator components, e.g., double-, zero-quantum coherences, etc., in the Liouville operator space of the two spin system under the propagator $U_i(t_p, 0)$,

$$\rho(t_p) = U_i(t_p, 0)\rho(0)U_i(t_p, 0)^+ = \rho(0) - i\int_0^{t_p}[H_i(t'), \rho(0)]dt' + .... \qquad (31)$$

If the duration $t_p$ is much shorter than the total duration $T_p$, that is, $T_p \gg t_p$, the created operator components in Eq.(31) is small and can be neglected in the contribution of the total propagator $U_i(T_p)$. Here the norm of the interaction Hamiltonian $H_i(t)$ keeps unchanged during the whole duration $T_p$, as can be seen from Eqs.(3) and (8). Therefore, the total propagator $U_i(T_p)$ may be approximated by $U_i(T_p, t_p)$ within the region SW($t_p$) when $T_p \gg t_p$. Since the interaction Hamiltonian $H_i(t)$ can be approximated as the longitudinal two-spin order operator ($2I_zS_z$) within the region SW($t_p$) when $t > t_p$, it follows from Eq.(29) that when $T_p \gg t_p$ the total propagator within the inversion plane region SW($t_p$) is given approximately by

$$U_i(T_p) \approx U(T_p, t_p) \approx \exp(-i\pi J 2 I_z S_z T_p). \tag{32}$$

It is well known that such a propagator can not derive the Hartmann-Hahn transfer. On the other hand, outside the resonance offset plane region SW($T_p$), that is,

$$\Delta\omega_q \geq W_q(T_p) + \delta W_q(T_p) \text{ and } \Delta\omega_q \leq -(W'_q(T_p) + \delta W'_q(T_p)) \ (q = i, s),$$

the adiabatic inversion pulses do not take into action, and the propagators are given by $U_i(T_p) = \exp(-i\pi J 2 I_z S_z T_p)$ and $U_0(T_p) = \exp[-i(\Omega_i I_z + \Omega_s S_z)T_p]$, respectively, as can be seen from Eqs.(1)-(3), since outside the region SW($T_p$) the coefficients $\alpha_q(\Delta\omega_q, T_p) = 1$, $\beta_q(\Delta\omega_q, T_p) = 0$, and $\gamma_q(\Delta\omega_q, T_p) = 0$ in Eqs.(9a) and (9b). The propagators also can not derive the Hartmann-Hahn transfer. For the transition region between the resonance offset plane regions SW($T_p$) and SW($t_p$) the interaction Hamiltonian is complicated and is given generally by Eq.(10). It is usually difficult to derive broadband Hartmann-Hahn transfer by the Hamiltonian. Therefore, a pair of adiabatic inversion pulses applied simultaneously to two heteronuclear coupled spins usually can not achieve effectively Hartmann-Hahn transfer between the two spins.

Numerical simulation is carried out in a heteronuclear two-spin system with scalar coupling constant $J=140$Hz under the full hyperbolic secant adiabatic inversion pulses with amplitude-modulation function $\omega_1(t) = \omega_0 \text{sech}(\beta t)$ ($-T_p/2 \leq t \leq T_p/2$) and frequency-modulation function constructed by Eq.(23) with the constant adiabatic factor $p=1/3$. There is difference between the backward-half and the full hyperbolic secant pulses, but the conclusion obtained from the numerical calculation below is the same for the two adiabatic pulses. For the case of the full hyperbolic secant pulse numerical simulation shows that during the beginning $T_p/4$ period of the pulse the longitudinal magnetization $I_z$ (or $S_z$) keeps unchanged over the inversion band, then the interaction Hamiltonian during the beginning $T_p/4$ period is the longitudinal two-spin order operator of Eq.(11) and its corresponding propagator is $U_i(T_p/4) = \exp(-i\pi J 2 I_z S_z T_p/4)$, and after the time $T_p/4$ the propagator is approximated by Eq.(32). Assume that the initial magnetization is $\rho(0) = I_x$. The total propagator (the pulse sequence) to be simulated numerically is taken as

$$R_T(t) = U_i(t) = U_0(t)^+ U_T(t), \tag{33a}$$

or $\quad R_T(t) = U_T(t) U_0(\phi_q(t) + \pi, t) \tag{33b}$

where the propagators $U_T(t)$ and $U_0(t)$ are given by Eq.(5) and (6) and their corresponding Hamiltonians are given by Eq.(1) and (2), respectively. The propagator $U_0(\phi_q(t)+\pi,t)$ with phase shift $+\pi$ with respect to the propagator $U_0(t)$ is used to refocus $U_0(t)$ approximately because here the pulses are 180 degree pulses. However, it must be pointed out that the propagator $U_0(t)$ usually can not be refocused by $U_0(\phi_q(t)+\pi,t)$ for non-inversion pulses. The density operator is calculated numerically according to $\rho(t_p) = R_T(t_p)I_x R_T(t_p)^+$ with the propagator (33a) or (33b). Figure 4 shows the resonance offset dependence for the anti-phase magnetization ($2I_y S_z$) created by the propagator $R_T(T_p)$ of Eq.(33b) from the initial magnetization $I_x$ on the resonance offset plane $(\Delta\omega_i, \Delta\omega_s)$: $|\Delta\omega_q|/2\pi \leq 20000 Hz$ ($q = i, s$) when the pulse duration $T_p=3/2J$. It can be seen that the initial magnetization $I_x$ is almost completely transferred into the anti-phase magnetization $-(2I_y S_z)$ on the resonance offset plane region, while the theoretic propagator $U_i(T_p)$ of Eq.(32) also predicts that initial magnetization $I_x$ should be completely transferred into the anti-phase magnetization $-(2I_y S_z)$ when the duration $T_p=3/2J$. Numerical simulation also shows that the initial magnetization $I_x$ is almost completely transferred into the magnetization $-I_x$ when $T_p=1/J$, which is also consistent with the prediction by the propagator of Eq.(32). These numerical simulations using the propagator (33b) and the propagator $U_i(t)$ of Eq.(33a) of the interaction Hamiltonian $H_i(t)$ show that it is difficult to achieve effectively broadband Hartmann-Hahn transfer in the two-spin system under the hyperbolic secant adiabatic inversion pulses. Therefore, a pair of adiabatic inversion pulses applied simultaneously to two heteronuclear coupled spins usually can not derive effectively Hartmann-Hahn transfer between the two spins. Actually, It is first understanding this point that leads me to the new idea to build up heteronuclear Hartmann-Hahn transfer sequences with the quasi-adiabatic excitation (90 degree) pulses.

**(b) the quasi-adiabatic excitation (90 degree) pulses**
Now consider the case that the pair of pulses used to derive the Hartmann-Hahn transfer in a heteronuclear two-spin system are taken as two quasi-adiabatic excitation (90 degree) pulses instead of the adiabatic inversion pulses. Then the conversion coefficients $\alpha_q(\Delta\omega_q, t_p)$ ($q = i, s$), etc., in Eqs.(9a) and (9b) satisfy the following conditions just like those conditions (24)-(26) of an adiabatic inversion pulse.
(1) within the excitation band $\Delta W_q(t_p)$, i.e., $-W'_q(t_p) \leq \Delta\omega_q \leq W_q(t_p)$, as shown in the schematic Figure 3 (A), the conversion coefficients fulfill
$\alpha_q(\Delta\omega_q, t_p) = 0$, $\beta_q(\Delta\omega_q, t_p) \neq 0$, and $\gamma_q(\Delta\omega_q, t_p) \neq 0$.

These shows that under the quasi-adiabatic excitation pulses the initial longitudinal magnetization $M_0$ ($I_z$ or $S_z$) is converted completely into the transverse magnetization $M_x$ and $M_y$ within the excitation band.

(2) in the transition regions: $W_q(t_p) < \Delta\omega_q < W_q(t_p) + \delta W_q(t_p)$ and $-W'_q(t_p) > \Delta\omega_q > -(W'_q(t_p) + \delta W'_q(t_p))$, the initial longitudinal magnetization $M_0$ is converted partly into the transverse magnetization $M_x$ and $M_y$,

$\alpha_q(\Delta\omega_q, t_p) \neq 0$, $\beta_q(\Delta\omega_q, t_p) \neq 0$, and $\gamma_q(\Delta\omega_q, t_p) \neq 0$.

(3) in the large resonance offset ranges, i.e., $\Delta\omega_q \geq W_q(t_p) + \delta W_q(t_p)$ and $\Delta\omega_q \leq -(W'_q(t_p) + \delta W'_q(t_p))$, the initial longitudinal magnetization $M_0$ keeps unchanged, that is,

$\alpha_q(\Delta\omega_q, t_p) = 1$, $\beta_q(\Delta\omega_q, t_p) = 0$, and $\gamma_q(\Delta\omega_q, t_p) = 0$.

Obviously, the quasi-adiabatic 90 degree pulses do not take into action outside their excitation band.

The quasi-adiabatic excitation (90 degree) pulses have the same property as the adiabatic inversion pulses, that is, flip angle of the pulses is independent of the pulse duration and excitation bandwidth of the pulses increases as the pulse duration and the square of the RF power. Therefore, the excitation bandwidth $\Delta W_q(t)$ at the duration $t$ is wider than $\Delta W_q(t_p)$ at the duration $t_p$ if $t > t_p$. By using the property one may find the possible reason why it could be possible for the quasi-adiabatic 90 degree pulses to derive the heteronuclear Hartmann-Hahn transfer. Now in the resonance offset plane region $SW(t_p)$: $-W'_q(t_p) \leq \Delta\omega_q \leq W_q(t_p)$ ($q = i, s$), as shown in Figure 3 (B), the interaction Hamiltonian $H_i(t)$ of Eq.(10) can be simplified by the excitation condition of the quasi-adiabatic 90 degree pulses, that is, $\alpha_q(\Delta\omega_q, t) = 0$, when any duration $t$ of the Hartmann-Hahn transfer sequence is longer than $t_p$, i.e., $t > t_p$. In this case the interaction Hamiltonian $H_i(t)$ is reduced to the form of Eq.(12), indicating that the interaction Hamiltonian now is an even-order multiple-quantum operator in the region $SW(t_p)$ instead of the longitudinal two-spin order operator which is the interaction Hamiltonian in the case of the adiabatic inversion pulses. When the total duration $T_p$ of the Hartmann-Hahn sequence is much longer than the time $t_p$, that is, $T_p \gg t_p$ the even-order multiple-quantum interaction Hamiltonian makes the main contribution to the whole Hartmann-Hahn transfer in the region $SW(t_p)$. It is shown in the former sections that an even-order multiple-quantum Hamiltonian may derive the Hartmann-Hahn transfer. Then the Hartmann-Hahn transfer may be achieved within the resonance offset plane region $SW(t_p)$ by the pair of the quasi-adiabatic 90 degree pulses. This is completely different from that case of the adiabatic inversion pulses. Therefore, it could be possible for a pair of quasi-adiabatic excitation (90 degree) pulses applied simultaneously to two heteronuclear coupled spins to derive the Hartmann-Hahn transfer between the two spins.

Numerical simulation is used to investigate the Hartmann-Hahn transfer process derived by the hyperbolic secant quasi-adiabatic 90 degree pulses in a two-spin system with J=140Hz. The frequency-modulation function of the quasi-adiabatic pulse is generated by Eq.(23) with the constant adiabatic factor $p=2.3$. The propagators (pulse sequences) used in numerical simulation include the propagator $U_i(t)$ of Eq.(33a) of the interaction Hamiltonian $H_i(t)$ and the propagator (33b). Figure 5 shows that the initial longitudinal magnetization $\rho(0) = I_z$ of the spin I can be transferred to the transverse magnetization $M_{xy}$ (absolute value $M_{xy} = \sqrt{M_x^2 + M_y^2}$) of the spin S under the propagator of Eq.(33b) of the full hyperbolic secant quasi-adiabatic 90 degree pulse in the resonance offset plane region: $|\Delta\omega_q|/2\pi \leq 20000$Hz ($q = i, s$). One can see that the Hartmann-Hahn transfer completely from the spin I to the spin S may occur even at some large resonance offset of the spin S, although the transfer is not uniform and may not be achieved completely at some resonance offset of the spin I. This shows that the hyperbolic secant quasi-adiabatic excitation pulse could provide a possibility to achieve the heteronuclear Hartmann-Hahn transfer. However, according to the property of even-order multiple-quantum operators [28, 33, 34] an even-order multiple-quantum Hamiltonian generated by the hyperbolic secant quasi-adiabatic pulses should not derive the transfer from the initial longitudinal magnetization $\rho(0) = I_z$ to any transverse magnetization such as $M_{xy}$ of the spin S. The reason why the numerical simulation shows the transfer $\rho(0) = I_z \rightarrow M_{xy}$ induced by the propagator of Eq.(33b) is that the propagator of Eq.(33b) is not the pure even-order multiple-quantum propagator $U_i(t)$ since the propagator $U_0(t)$ usually can not be refocused by $U_0(\phi_q(t) + \pi, t)$. In fact, the numerical simulation further shows that the initial longitudinal magnetization $\rho(0) = I_z$ is almost completely transferred to the double- and zero-quantum coherences by the even-order multiple-quantum propagator of Eq.(33a): $U_i(t) = U_0(t)^+ U_T(t)$, where the propagator $U_0(t)$ of the propagator $U_T(t)$ of Eq.(5) is refocused completely by $U_0(t)^+$. The backward-half hyperbolic secant quasi-adiabatic 90 degree pulse is used to investigate the transfer in the numerical calculation. Figure 6 shows that the initial longitudinal magnetization $\rho(0) = I_z$ in the two-spin system with J=140Hz is almost completely transferred to the double- and zero-quantum coherences in the resonance offset plane region: $20000 \text{Hz} < \Delta\omega_q / 2\pi < 90000 \text{Hz}$ ($q = i, s$) under the propagator $U_i(t_p)$ of the backward-half hyperbolic secant quasi-adiabatic 90 degree pulses. The complete transfer occurs at the pulse duration of about $t_p \approx 1/2J$. These simulations confirm the theoretic analysis that the interaction Hamiltonian $H_i(t)$ in the resonance offset region is approximately an even-order multiple-quantum operator.

## 4. A possible ultra-broadband heteronuclear Hartmann-Hahn sequence

Although it could be possible for the quasi-adiabatic 90 degree pulses to derive heteronuclear Hartmann-Hahn transfer, it is usually difficult to obtain broadband heteronuclear Hartmann-Hahn transfer by a simple sequence of the quasi-adiabatic 90 degree pulses. The multiple-quantum operator algebra space formalism [28] may be very helpful in the construction of an ultra-broadband heteronuclear Hartmann-Hahn sequence with the quasi-adiabatic 90 degree pulses. According to the theoretic analysis in the former sections the heteronuclear Hartmann-Hahn transfer can be performed in the even-order multiple-quantum operator space. This principle may guide one to design ultra-broadband heteronuclear Hartmann-Hahn sequences by using the quasi-adiabatic 90 degree pulses.

A possible ultra-broadband heteronuclear Hartmann-Hahn transfer sequence could be constructed with the quasi-adiabatic 90 degree pulses on the basis of the theoretic analysis in the even order multiple-quantum operator subspace. Figure 7 (A) shows the Hartmann-Hahn transfer sequence. The initial longitudinal magnetization $I_z$ of the spin $I$ is first transferred completely into the even-order multiple-quantum coherence operators by the first even-order multiple-quantum unitary propagator $U_{e1}(t_p)$, then under the second even-order multiple-quantum unitary propagator $U_{e2}(t_p)$ the even-order multiple-quantum coherence operators are transferred completely to the longitudinal magnetization $S_z$ of the spin $S$. The sequence consists of the quasi-adiabatic 90 degree pulses which are used to prepare the two even-order multiple-quantum propagators. This sequence is called quasi-adiabatic excitation pulse deriving ultra-broadband Heteronuclear Hartmann-Hahn polarization transfer echo, as shown in the schematic picture of Figure 7 (B). The polarization transfer echo is achieved in the even-order multiple-quantum operator subspace. The propagator $U_0(t)$ of the quasi-adiabatic 90 degree pulses applied simultaneously to two non-interacting spins have not net contribution to the transfer, but it can degrade the echo and even destroy the echo and hence need to be refocused effectively. How to refocus effectively the propagator $U_0(t)$ is a challenge in implementing the Hartmann-Hahn transfer echo experiment at present. Now one possible Hartmann-Hahn transfer echo sequence could be constructed below. The first even-order multiple-quantum unitary propagator $U_{e1}(t_p)$ is prepared by a pair of the backward-half hyperbolic secant quasi-adiabatic 90 degree pulses applied simultaneously to the two spin with the RF phase along X-direction and a refocusing propagator $U_0(t_p)^+$,

$$U_{e1}(t_p)_X = U_i(t_p)_X = (U_0(t_p)^+ U_T(t_p))_X.$$

The second even-order multiple-quantum unitary propagator $U_{e2}(t_p)$ is prepared by the pair of backward-half hyperbolic secant quasi-adiabatic 90 degree pulses applied simultaneously to the two spin with the RF phase along Y-direction and a refocusing propagator $U_0(t_p)^+$

$$U_{e2}(t_p)_Y = U_i(t_p)_Y = (U_0(t_p)^+ U_T(t_p))_Y.$$

The numerical simulation has been carried out for the echo sequence of Figure 7 (A) which has the propagator $U_{e2}(t_p)_Y U_{e1}(t_p)_X$. Figure 8 shows that near complete Hartmann-Hahn transfer ($I_z \to S_z$) is achieved by the echo sequence using the backward-half hyperbolic secant quasi-adiabatic 90 degree pulses whose frequency-modulation functions are constructed with the approximated method of Eq.(23) with the constant adiabatic constant *p=2.3*. It can be seen from Figure 8 that a more than 80KHz bandwidth ($20000 Hz < \Delta\omega_q / 2\pi < 100000 Hz$ (*q = i, s*)) for the Hartmann-Hahn transfer (the transfer efficiency > 90% ) in each of the two resonance offsets ($\Delta\omega_i$, $\Delta\omega_s$) is obtained by using only 5KHz of the RF peak power of the pulses. The numerical simulation also shows that the Hartmann-Hahn transfer based on this echo sequence is uniform over the resonance offset plane: $20000 Hz < \Delta\omega_q / 2\pi < 100000 Hz$ (*q = i, s*). These results illustrate for the first time the possibility that an ultra-broadband heteronuclear Hartmann-Hahn transfer could be achieved by the quasi-adiabatic excitation pulses with a lower RF power just similar to that an ultra-broadband inversion can be obtained by an adiabatic inversion pulse with a lower RF power [14, 15]. These results also show that the multiple-quantum operator algebra space formalism [28] is very helpful for the construction of the ultra-broadband heteronuclear Hartmann-Hahn transfer echo sequence.

From the view point of the quasi-adiabatic pulses a longer pulse duration $t_p$ corresponds to a wider excitation band of the pulses, which results in also a wider excitation band of the even-order multiple-quantum interaction Hamiltonian and hence a more broadband Hartmann-Hahn transfer. On the other hand, the complete Hartmann-Hahn transfer is also dependent on the scalar coupling between the two heteronulcear nuclei and usually is approximately proportional to inverse scalar coupling constant [5, 6]. It seems that the bandwidth for the Hartmann-Hahn transfer is limited by the scalar coupling, but the excitation band for the quasi-adiabatic excitation pulses is dependent on not only the pulse duration but also the square of the RF power [27]. In particular, bandwidth of the quasi-adiabatic pulses increases in a quadratic form as the RF power, indicating that the bandwidth of the Hartmann-Hahn transfer sequence should increase as the square of the RF power. This is the extraordinary performance of the heteronuclear Hartmann-Hahn transfer sequences based on the quasi-adiabatic 90 degree pulses. It indicates that ultra-broadband heteronuclear Hartmann-Hahn transfer could be achieved by the quasi-adiabatic 90 degree pulses with low RF power.

Finally it must be pointed out that the key to implement in NMR experiments the ultra-broadband heteronuclear Hartmann-Hahn transfer echo should be how the propagator $U_0(t)$ of the quasi-adiabatic 90 degree pulses can be refocused effectively.

**Acknowledgement**

The work was supported by NIH grant P41RR08079 of the Center for Magnetic Resonance Research, Minnesota university medical school. Author thanks Professor Michael Garwood and Dr. Rolf Gruetter for their helps.**References** and usually is approximately proportional to inverse scalar coupling constant [5, 6]. It seems that the bandwidth for the Hartmann-Hahn transfer is limited by the scalar coupling, but the excitation band for the quasi-adiabatic excitation pulses is dependent on not only the pulse duration but also the square of the RF power [27]. In particular, bandwidth of the quasi-adiabatic pulses increases in a quadratic form as the RF power, indicating that the bandwidth of the Hartmann-Hahn transfer sequence should increase as the square of the RF power. This is the extraordinary performance of the heteronuclear Hartmann-Hahn transfer sequences based on the quasi-adiabatic 90 degree pulses. It indicates that ultra-broadband heteronuclear Hartmann-Hahn transfer could be achieved by the quasi-adiabatic 90 degree pulses with low RF power.

Finally it must be pointed out that the key to implement in NMR experiments the ultra-broadband heteronuclear Hartmann-Hahn transfer echo should be how the propagator $U_0(t)$ of the quasi-adiabatic 90 degree pulses can be refocused effectively.

**Acknowledgement**

The work was supported by NIH grant P41RR08079 of the Center for Magnetic Resonance Research, Minnesota university medical school. Author thanks Professor Michael Garwood and Dr. Rolf Gruetter for their helps.

**References**

1. S.R.Hartmann and E.L.Hahn, Phys.Rev. 128, 2042 (1962)
2. L.Muller and R.R.Ernst, Mol.Phys. 38, 963 (1979)
3. G.C.Chingas, A.N.Garroway, R.D.Bertrand, and W.B.Moniz, J.Chem.Phys. 74, 127 (1981)
4. M.Mehring, "High Resolution NMR spectroscopy in solids," 2nd Edn, Spring-Verlag, Berlin, 1983
5. R.R.Ernst, G.Bodenhausen, and A.Wokaun, "Principle of Nuclear Magnetic Resonance in One and Two Dimensions," Clarendon Press, Oxford, 1987
6. S.J.Glaser and J.J.Quant, Adv.Magn.Reson. 19, 59 (1996)
7. R.Freeman, "Spin Choreography," Spektrum, Oxford, 1997
8. A.J.Shaka, C.J.Lee, and A.Pines, J.Magn.Reson. 77, 274 (1988)
9. N.Sunitha Bai, N.Hari, and R.Ramachandran, J.Magn.Reson. A 106, 248 (1994)
10. M.G.Schwendinger, J.Quant, J.Schleucher, S.J.Glaser, and C.Griesinger, J.Magn.Reson. A111, 115 (1994)
11. T.Carlomagno, B.Luy, and S.J.Glaser, J.Magn.Reson. 126, 110 (1997)
12. T.Carlomagno, M.Maurer, M.Sattler, M.G.Schwendinger, S.J.Glaser,

**Figure Captions**

Figure One: The excitation profiles of the backward-half hyperbolic secant quasi-adiabatic excitation (90 degree) pulse $\omega_1(t_p) = \omega_0 \, \text{sech}(\beta t_p)$ ($t_p \geq 0$) with different pulse duration ($t_p$). The frequency-modulation function is generated by the conventional approximated formula (23) with the constant adiabatic factor *p=2.3*. In numerical simulation the pulse has the RF power $\omega_0/2\pi = 5\text{KHz}$, the truncation parameter $\beta = 5.9865/T_p$ ($T_p$=0.007s), and different pulse duration: (a) $t_p$=7ms, (b) $t_p$=7/2ms, (c) $t_p$=7/4ms, (d) $t_p$=7/8ms, (e) $t_p$=7/16ms. (A) the quasi-adiabatic 90 degree pulse with the unitary propagator $U(t_p)^+$, (B) the quasi-adiabatic 90 degree pulse with the unitary propagator $U(t_p)$.

Figure Two: The inversion profiles of the backward-half hyperbolic secant adiabatic inversion pulse with different pulse duration ($t_p$). The frequency-modulation function is generated by the approximated formula (23) with the constant adiabatic factor *p=1/3*. In numerical simulation the pulse has the RF power $\omega_0/2\pi = 5\text{KHz}$, the truncation parameter $\beta = 5.9865/T_p$ ($T_p$=0.007s), and different pulse duration: (a) $t_p$=7ms, (b) $t_p$=7/2ms, (c) $t_p$=7/4ms, (d) $t_p$=7/8ms, (e) $t_p$=7/16ms. (A) the adiabatic inversion pulse with the unitary propagator $U(t_p)^+$, (B) the adiabatic inversion pulse with the unitary propagator $U(t_p)$.

Figure Three: (A) the schematic inversion and excitation profiles of the adiabatic inversion pulses and the quasi-adiabatic excitation pulses showing how the flip angle of the pulses does not change and their bandwidth increases as the pulse duration. (B) the schematic resonance offset plane region on which the adiabatic and quasi-adiabatic pulses take into action.

Figure Four: The resonance offset profile of the polarization transfer from the initial transverse magnetization $I_x$ to the magnetization ($2I_yS_z$) under the full hyperbolic secant adiabatic inversion pulses applied to the two spins (IS) with scalar coupling constant J=140Hz. The pulses have the RF power $\omega_0/2\pi = 5\text{KHz}$, the truncation 1%, and the pulsewidth near 3/2J.

Figure Five: The resonance offset profile of the Hartmann-Hahn transfer from the I-spin longitudinal magnetization $I_z$ to the S-spin transverse magnetization ($M_{xy}$) under the full hyperbolic secant quasi-adiabatic 90 degree pulses applied to the two spins (IS) with J=140Hz. The propagator (pulse sequence) is $R_T(t)$ of Eq.(33b). The pulses have the RF power $\omega_0/2\pi = 5\text{KHz}$, the truncation 1%, and the pulsewidth near 1/J.

Figure Six: The resonance offset profile of the even-order multiple-quantum coherences created by the backward-half hyperbolic secant quasi-adiabatic 90 degree pulses from the I-spin longitudinal magnetization $I_z$ using the propagator $U_i(t_p) = U_0(t_p)^+ U_T(t_p)$ in the two-spin system (IS) with J=140Hz. The pulses have the RF power $\omega_0/2\pi = 5\text{KHz}$, the truncation parameter $\beta = 10.5966/t_p$, and the pulsewidth $t_p$=3.6ms (1/2J). The absolute value of the created total even-order multiple-quantum coherences is denoted as $\text{EVOMQ} = \sqrt{\sum_{p,q=x,y} \langle 2I_pS_q \rangle^2}$,

where $\langle 2I_p S_q \rangle$ is the coefficient of the created multiple-quantum coherence $2I_p S_q$. The shown highest contour level of EVOMQ is greater than 95%.

Figure Seven: (A) The ultra-broadband heteronuclear Hartmann-Hahn polarization transfer echo sequence. The two propagators $U_{en}(t_p)$ (n = 1, 2) are the even-order multiple-quantum propagators and could be prepared by the quasi-adiabatic 90 degree pulses. The transfer is from the longitudinal magnetization $I_z$ ($S_z$) to the longitudinal magnetization $S_z$ ($I_z$). (B) the schematic diagram of the quasi-adiabatic excitation pulse deriving heteronuclear Hartmann-Hahn polarization transfer echo. EVOMQOS: the even-order multiple-quantum operator subspace.

Figure Eight: The ultra-broadband heteronuclear Hartmann-Hahn transfer ($I_z \to S_z$) derived by the backward-half hyperbolic secant quasi-adiabatic 90 degree pulses using the quasi-adiabatic excitation pulse deriving heteronuclear Hartmann-Hahn polarization transfer echo sequence in Figure 7 (A). Each of the pulses has the RF power $\omega_0/2\pi = 5KHz$, the truncation parameter $\beta = 10.5966/t_p$, and the pulsewidth $t_p$=3.6ms (1/2J). The shown highest contour level of the magnetization $S_z$ is greater than 91%.

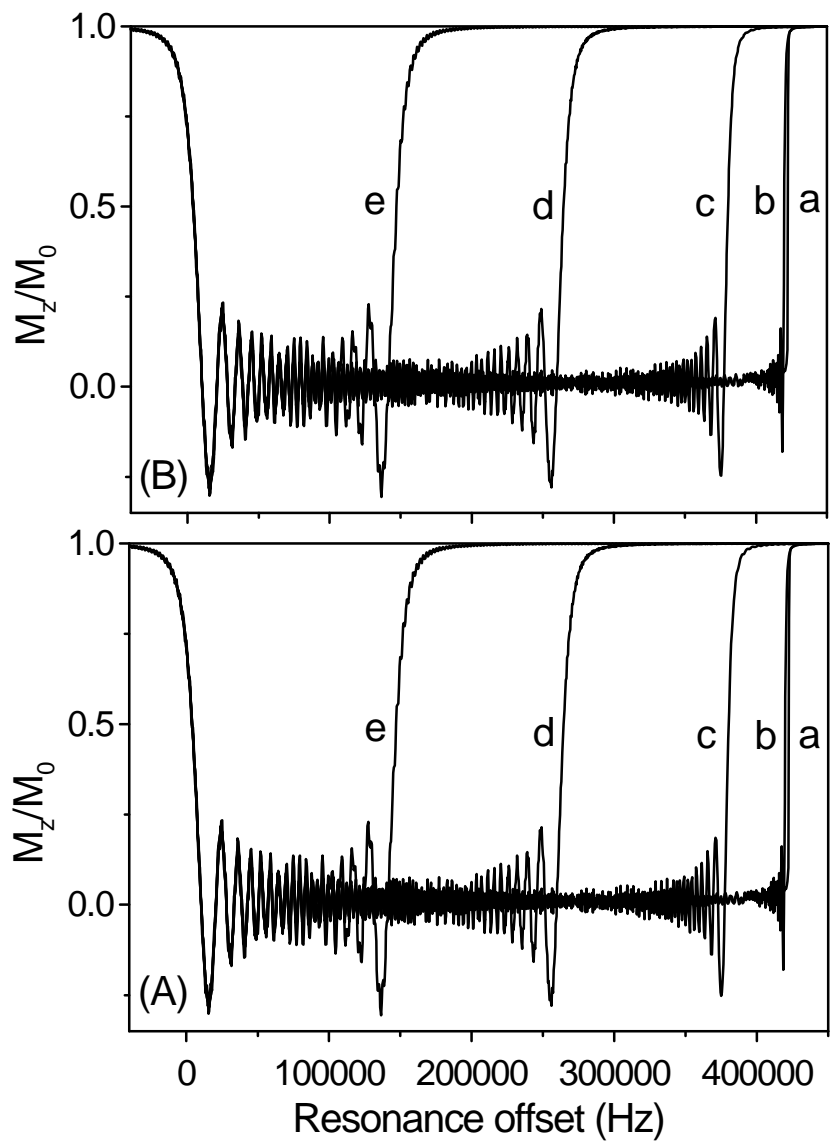

(Figure One)

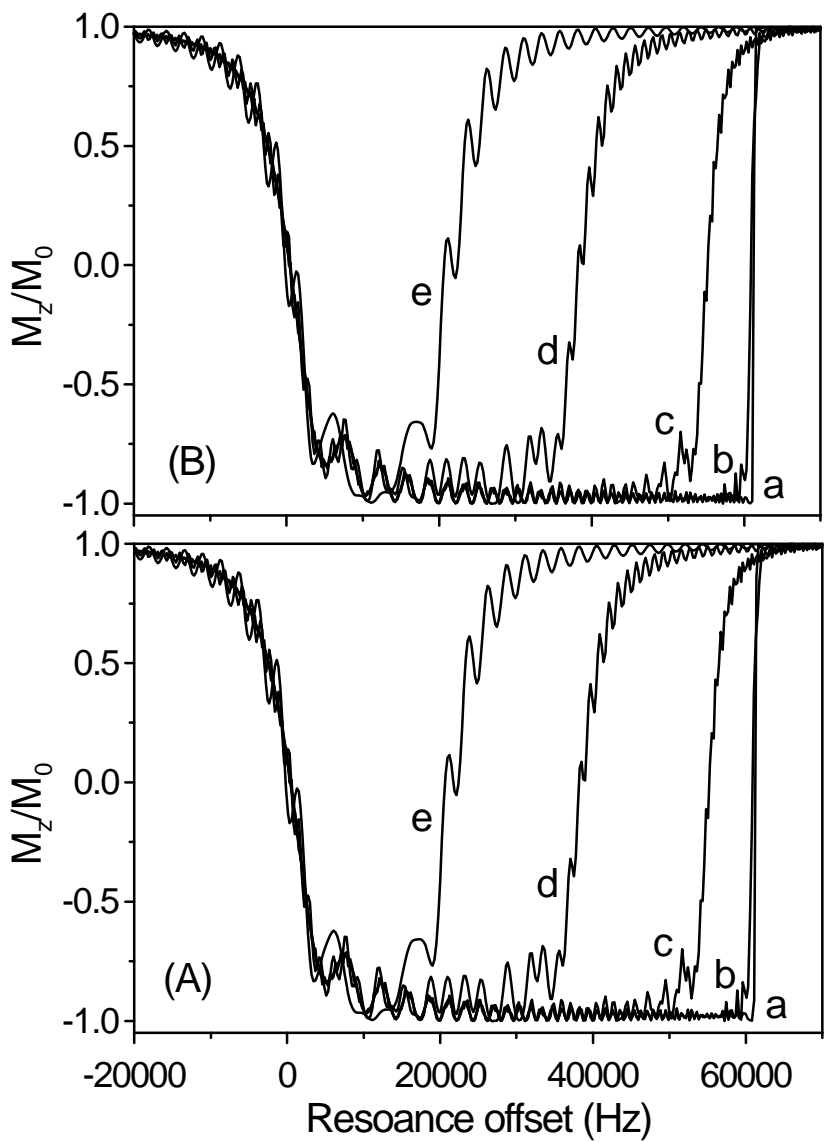

(Figure Two)

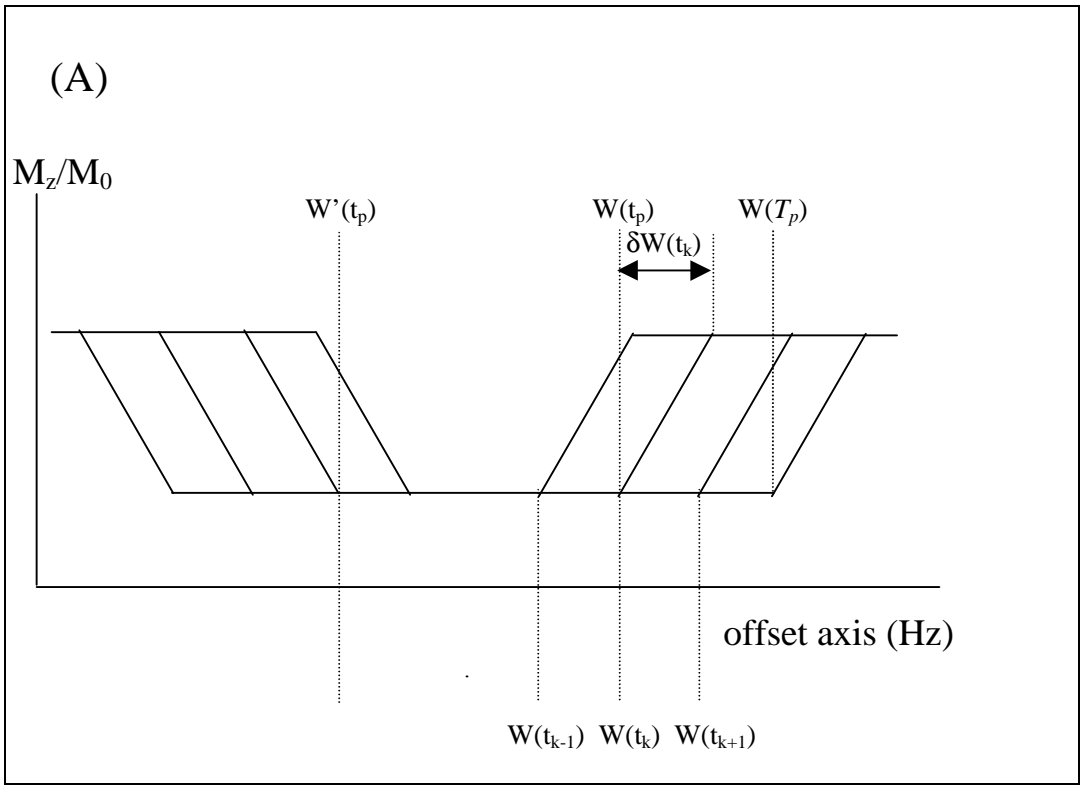

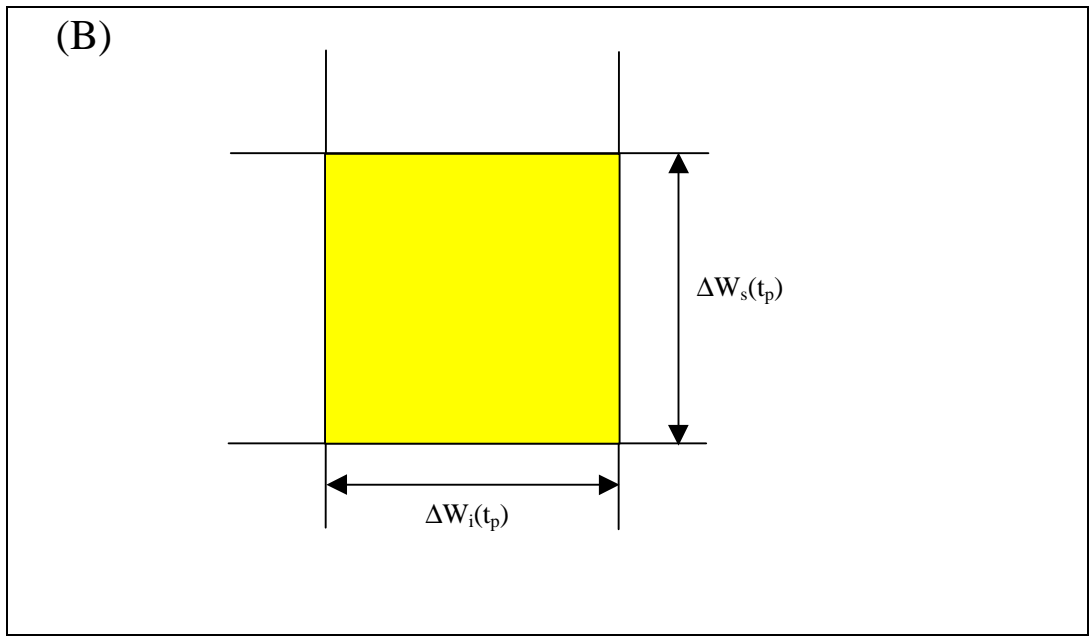

(Figure Three)

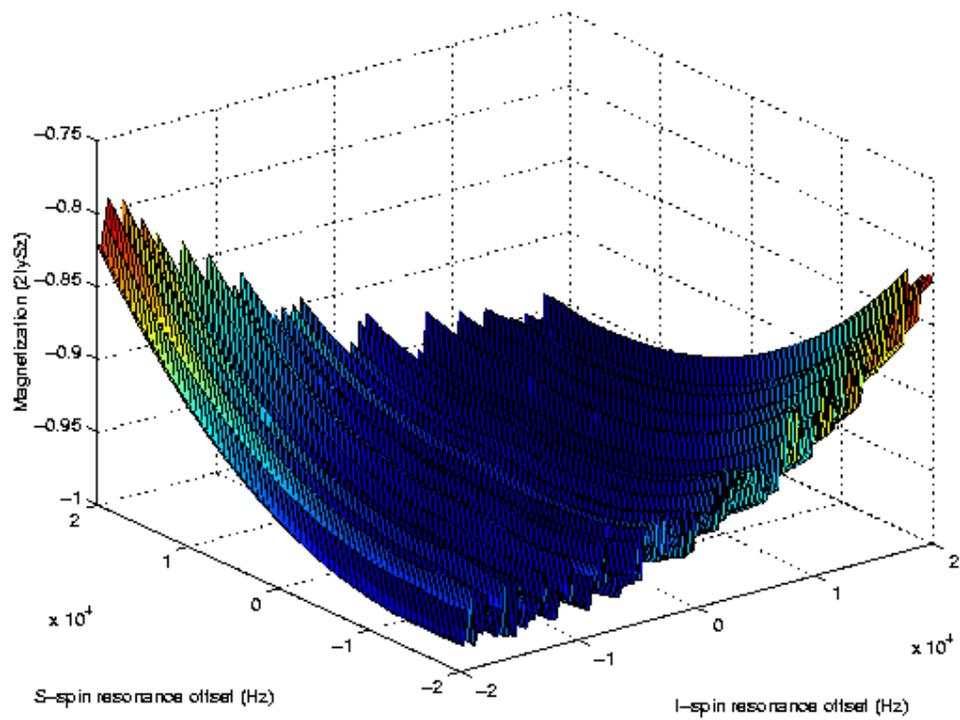

(Figure Four)

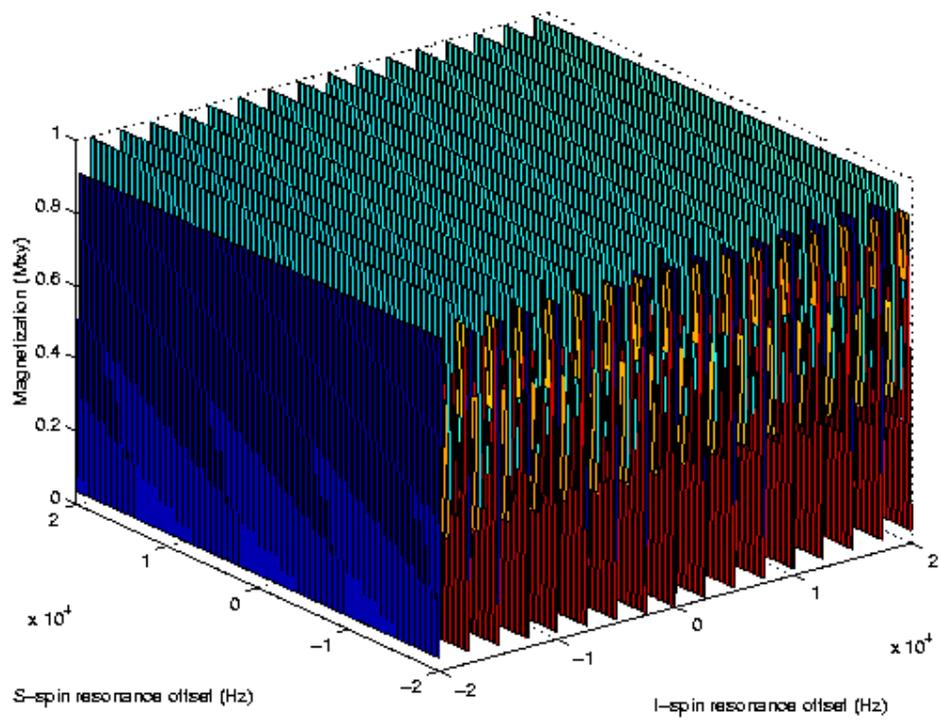

(Figure Five)

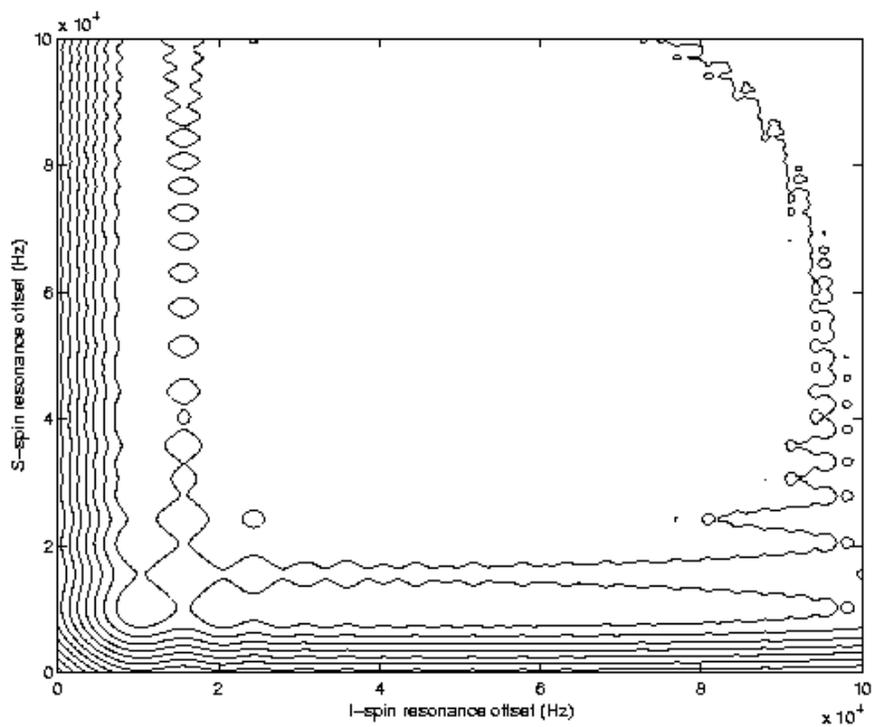

(Figure Six)

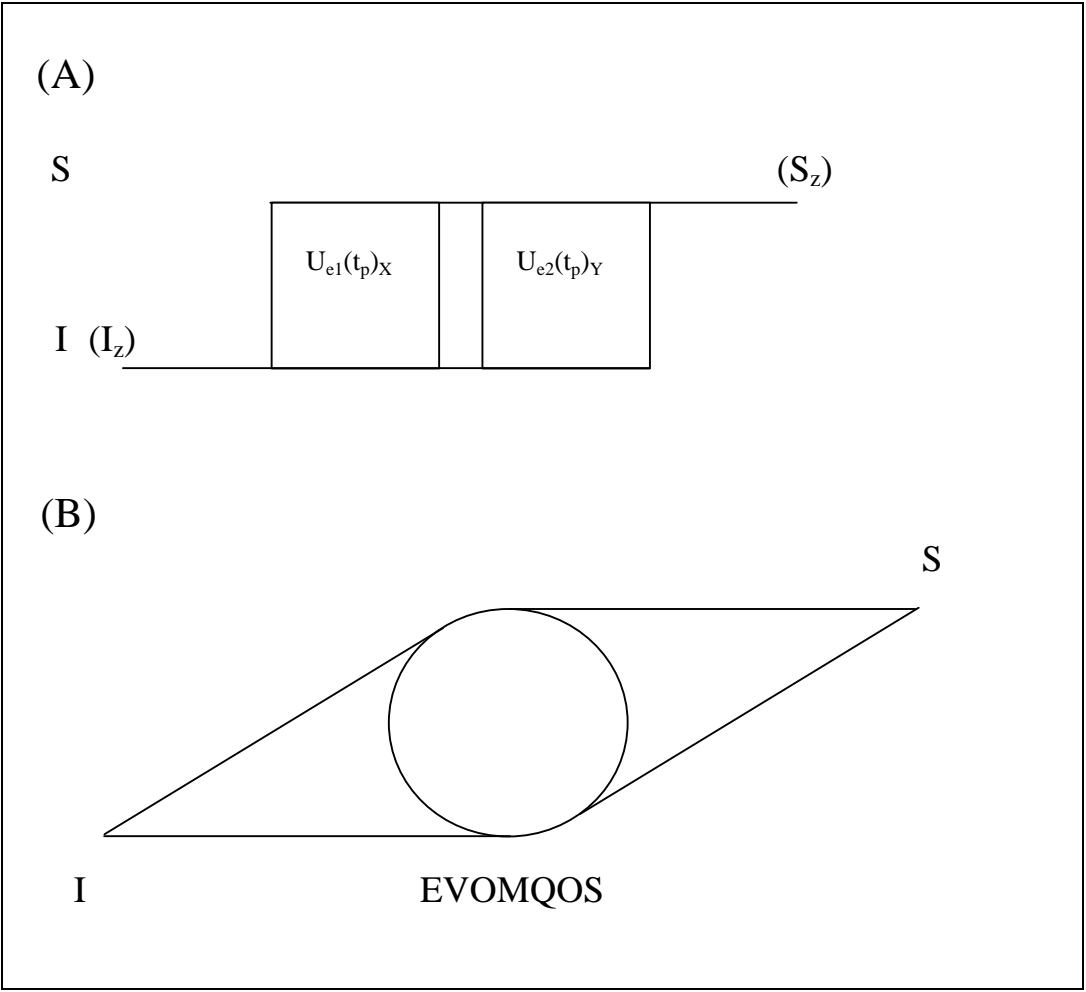

(Figure Seven)

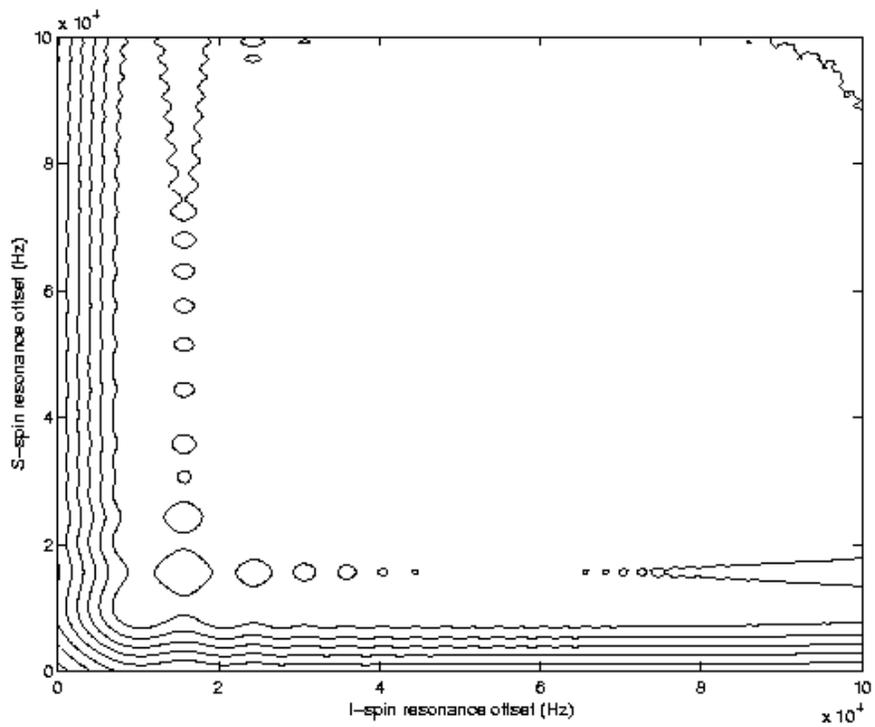

(Figure Eight)